\documentstyle[twoside,fleqn,espcrc2]{article}

\newcommand{\AmS}{{\protect\the\textfont2
  A\kern-.1667em\lower.5ex\hbox{M}\kern-.125emS}}

\title{Prospects for perfect actions}

\author{P. Hasenfratz \address{Institute for Theoretical Physics, 
        University of Bern, \\ 
        Sidlerstr. 5, CH-3012 Bern, Switzerland}%
        \thanks{Work supported in part by Schweizerischer Nationalfonds
                and by Iberdrola, Ciencia y Tecnologia, Espana.}}
       
\begin{document}

\begin{abstract}
The fixed-point (FP) action in QCD, although it is local and determined
by classical equations, is difficult to parametrize well and is expensive
to simulate. But the stake is high: the FP action has scale invariant 
instanton solutions, has no topological artifacts, satisfies the index
theorem on the lattice, does not allow exceptional configurations, requires
no tuning to get the pion massless and is expected to reduce the cut-off
effects significantly. An overview is given including a discussion
on tests in Yang-Mills theory, QCD and $d=2$ spin and gauge models.
\end{abstract}

\maketitle

\section{INTRODUCTION}

This is a review on the most important properties of perfect lattice
actions and on the present status of constructing, parametrizing and
simulating them. This summary relies on earlier lattice conference
presentations \cite{lat}, on published results and on contributions to this
conference, adding some new developments also which are not published yet.

I will call an action classically perfect if its classical predictions
on the lattice (fine, or coarse) agree with those in the continuum.
The quantum perfect action does the same in the quantum theory. These 
actions have beautiful properties which one would not expect a
lattice action can have.

The actions we consider are local. Locality is a basic requirement for
any action to keep renormalizability and universality. A local regularized
action has a space-time extension of $O(1/\Lambda^{cut})$. 

Although these actions are local, they contain infinitely many coupling
constants. These couplings are not free, but are fixed by classical
equations, 
or by path integrals. The main problem is not so much to determine
these couplings, but to parametrize these actions in terms of a finite 
number of couplings so that a numerical simulation remains feasable and
the important properties are preserved to a good extent.

The future of this approach depends also on the success/failure of other,
more traditional methods to overcome the main problems of lattice
calculations. Here I do not only mean the elimination of cut-off effects,
but also the difficulties related to chiral symmetry, exceptional 
configurations and topology (instanton solutions and index theorem) to
which the classically perfect actions offer a solution.

\section{THE FIXED-POINT EQUATIONS}

The notion of perfect lattice regularized actions is related to
Wilson's renormalization group (RG) theory \cite{wrg}. I will consider
asymptotically free $d=4$ and $d=2$ models. Consider a RG 
transformation (RGT) in QCD
\begin{equation}
\exp \{-[\beta'S'_g(V)+S'_f(\bar{\chi},\chi,V)] \}=
\label{rgt}
\end{equation}
\begin{displaymath}
\int D\bar{\psi}D\psi DU \exp \{-\beta [S_g(U)+\kappa_gT_g(V,U)]
\end{displaymath}
\begin{displaymath}
-[S_f(\bar{\psi},\psi,U)+\kappa_fT_f(\bar{\chi},\chi;\bar{\psi},\psi,U)] \}.
\end{displaymath}
Before the transformation, $\bar{\psi},\psi,U$ and $\beta S_g+S_f$ 
denote the quark and gauge fields and the local QCD action, respectively.
The corresponding fields on the blocked  (coarse)
lattice are $\bar{\chi},\chi$ and $V$. 
The averaging process is defined by $T$. The action and $T$ depend on 
parameters,
out of which only two are indicated explicitely: $1/\beta \sim g$ is the
AF coupling and $\kappa$ specifies the stiffness of the RG averaging. A basic
assumption of the RG theory is that the path integral on the r.h.s. of 
eq.~(\ref{rgt}) is a non-critical problem with short range fluctuations only
(due to the presence of the external fields $\bar{\chi},\chi,V$) and so,
the action on the l.h.s. is expected to be local also.

Consider first a YM theory. The flow diagram (giving the change of couplings 
in the action) is expected to be simple: there is a fixed-point (FP) $S^{FP}$ in
the $g=0$ hyperplane which has only one (weakly) relevant direction (associated
with the AF coupling $g$) running out of the plane and defining the renormalized
trajectory. It is easy to show that the points of the renormalized trajectory
define quantum perfect actions \cite{hn}.

Consider the FP, $S^{FP}$ (which lies in the $\beta = \infty$ plane), multiply
it by $\beta$: $\beta S^{FP}$ and allow $\beta$ to move away from $\infty$.
This is not a RG flow, but defines an action for every value of $\beta$. I
will call this action the FP action which is, as will be discussed later,
a classically perfect action.

The form of $S^{FP}$ is determined by classical equations \cite{hn}. 
For $\beta \rightarrow
\infty$ we can use the saddle-point approximation to calculate the r.h.s of
eq.~(\ref{rgt}). For the YM theory one obtains \cite{sol}
\begin{equation}
S_g^{FP}(V) = \min_{\{ U \}}(S_g^{FP}(U) + \kappa_g T_g(V,U)).
\label{fpg}
\end{equation}
Add now the fermions. Write the fermion
action and the block transformation in the form:
\begin{eqnarray}
\lefteqn{S_f(\bar{\psi},\psi,U) = \sum_{n,n'} \bar{\psi}_n 
h_{n,n'}(U) \psi_{n'}}, \\
\lefteqn{T_f(\bar{\chi},\chi;\bar{\psi},\psi,U)=}  \\
& & \sum_{n_B} (\bar{\chi}_{n_B}-\sum_n \bar{\psi}_n 
\omega(U)^+_{n,n_B})  \nonumber \\
& & (\chi_{n_B}-\sum_n \omega(U)_{n_B,n} \psi_n) ,\nonumber
\end{eqnarray} 
where $\omega(U)$ defines a gauge invariant averaging. This   
averaging function is defined to be diagonal in Dirac space. 
In the saddle-point 
approximation the integral over $U$ in eq.~(\ref{rgt})
is dominated by the minimizing
configuration $U = U(V)$ defined by eq.~(\ref{fpg}), while the remaining 
fermion integral is Gaussian. It follows then that the blocked action is quadratic
in the fermion fields also. Performing the Gaussian integral one obtains
a classical equation for $h^{FP}$ \cite{hfp1,hfp2}
\begin{eqnarray}
\label{ver}
\lefteqn{h^{FP}(V)^{-1}_{n_B,n'_B} = 1/\kappa_f \,\delta_{n_B,n'_B} + } \\
& & \sum_{n,n'}\omega(U)_{n_B,n}h^{FP}(U)^{-1}_{n,n'} 
\omega(U)^+_{n',n'_B} \,  , \nonumber
\end{eqnarray}
where $U=U(V)$. Gaussian integrals are equivalent to minimization. It will
be useful to write eq.~(\ref{ver}) also as a minimization
\begin{eqnarray}
\label{verp}
\lefteqn{\sum_{n_B,n'_B}\bar{\chi}_{n_B} h^{FP}(V)_{n_B,n'_B} \chi_{n'_B}= } \\
& & \min_{\bar{\psi},\psi}[S_f(\bar{\psi},\psi,U)+\kappa_f
T_f(\bar{\chi},\chi;\bar{\psi},\psi,U)], \nonumber
\end{eqnarray}
where $U=U(V)$ and $\psi,\chi$ are c-number fields.

\section{PROPERTIES OF THE FP ACTION IN QCD}

I summarize the most important exact properties of the FP action in QCD.
In my talk at the conference I argued that at non-overlapping points the
Green's functions are chiral symmetric if calculated with the quantum perfect
action. I replaced this part by the more relevant discussion on the chiral
properties of the FP action, Sect.~3.5.

\subsection{Free spectrum}

The free gauge and fermion part of the FP action reproduce the exact
relativistic spectrum of the continuum.

\subsection{Instanton solutions and the topological charge}
The FP Yang-Mills action has scale invariant instanton 
solutions \cite{hn,sol}. A consistent
topological charge can be defined \cite{top1,top2}. 
There are no topological artifacts.

A simple consequence of eq.~(\ref{fpg}) is that if the configuration $V$
satisfies the FP Euler-Lagrange equations of motion, $\delta S^{FP}(V)/\delta V = 0$,
then the minimizing configuration $U(V)$ satisfies it also. The two solutions
have the same action. Observe, that U(V) lives on a lattice which is finer
than that of $V$ by the scale factor of the RGT. Iterating this results
to the continuum one finds that the FP action has scale invariant instanton solutions
and the value of the action is equal to that in the continuum. In addition, one 
can define the topological charge $Q(U^{(0)})$ of a 
configuration $U^{(0)}$ (solution, or not) by identifying
$Q(U^{(0)})$ with the well defined and unique topological charge of the smooth
configuration $U^{(n)}$, where $U^{(1)}$ is the minimizing configuration of $U^{(0)}$,
$U^{(2)}$ is that of $U^{(1)}$,... and $U^{(n)}$ is the minimizing configuration of
$U^{(n-1)}$, n is sufficiently large. With this definition the action of {\em any}
configuration is larger than $8 \pi^2 \mid Q \mid$, i.e.  
no topological artifacts exist.

\subsection{Fermionic zero modes and the index theorem}
Using the FP action, the index theorem on the zero modes of the Dirac operator
remains valid even on coarse configurations \cite{ind}.

Using the same derivation as in Sect.~3.2, one obtains from eq.~(\ref{verp}) that if 
$\chi_{n_B}$ is a solution of the FP Dirac equation over
the background field $V$: $h^{FP}(V) \chi = 0$, then the minimizing configuration 
$\psi$ on the r.h.s
of eq.~(\ref{verp}) is a solution of the FP Dirac equation over the background
field $U(V)$ on the fine lattice. Since eq.~(\ref{verp}) is quadratic in the
fermion fields, the minimum is unique, and so the opposite statement is true also:
if $\psi$ is a solution of the FP Dirac equation over $U(V)$, then 
$\chi_{n_B} = \sum_n \omega(U)_{n_B,n} \psi_n$ is a solution on the coarse lattice
over $V$. This result can be iterated up, towards finer lattices, or down.

Take now an $U^{(0)}$ configuration with topological charge $Q$. The sequence
of minimizing configurations $U^{(1)},U^{(2)},...$ (Sect.~3.2) have the same 
topological charge.
Going up on this sequence we get finally to a very smooth configuration on which the
Atiyah-Singer index theorem \cite{as} is applicable. 
The discussion in the previous 
paragraph and the observation that $\omega$ is trivial in Dirac space
imply then that over each gauge configuration of the sequence, the FP Dirac
equation has the same number of zero modes with the same helicities as in the
continuum. 

\subsection{Exceptional configurations}
A trivial consequence of Sect.~(3.3) is that there are no exceptional configurations,
i.e. configurations on which the Dirac operator has a lattice artefact zero mode
for which no associated zero mode exists in the continuum \cite{ind}.

\subsection{Chiral symmetry}
The FP action has no doublers and it is not explicitly chiral symmetric. Nevertheless,
it has no tuning problem: the pion mass is zero when the bare quark mass in the action
is zero. It is expected also that all the soft pion theorems are valid as in the
continuum.

The free fermion FP propagator can be obtained by solving eq.~(\ref{ver}) 
with $V=U=1$ \cite{wie,ku}.
The structure of the solution is easy to see if we consider a RGT with
a very large scale factor ('blocking out of the continuum', \cite{wb}) .
On the r.h.s of eq.~(\ref{ver}) we have a very fine lattice and the propagator can be
replaced by the continuum propagator. In momentum space we get
\begin{eqnarray}
\label{free}
\lefteqn{ 1/\tilde{h}^{FP}_{free}(q) = 1/\kappa_f+ }  \\
& & \sum_{l\in Z^d} \frac{ i\gamma^{\mu}(q_{\mu}+2\pi l_{\mu}) }{(q+2\pi l)^2}
\mid \tilde{\omega}(q+2\pi l)\mid^2 .  \nonumber
\end{eqnarray}
The sum over the integer vectors $l=(l_1,...,l_d)$ enters when $k$ in the continuum
propagator is expressed in terms of $q$ which is defined over the Brillouin zone.
The propagator, and so the action, is not chiral invariant. The chiral symmetry
breaking term $1/\kappa_f$ comes entirely from the block transformation $\kappa_f T$.
In configuration space the effect of chiral symmetry breaking in the propagator shows
up at $n=0$ only. For finite scale transformations (which have to be iterated to reach
the FP) the breaking term, which can be expressed always as a function of $\omega$, 
might become $q$ dependent, but remains analytic \cite{ku}. In general, we have
\begin{equation}
\label{gwf}
\{ 1/\tilde{h}^{FP}_{free}(q) ,\gamma^5 \} = 2 \gamma^5 \frac{1}{\kappa_f} \tilde{R}(q) .
\end{equation}
In configuration space, $R(n)$ is local, it is a smeared $\delta_{n,0}$. At 
$\kappa_f = \infty$ the action will be chiral symmetric, but, at the same time, becomes 
non-local \cite{wie,wb}, in consistency with the Nielsen-Ninomiya theorem \cite{nn}.

More than fifteen years ago, Ginsparg and Wilson \cite{gw}
suggested to elevate the free field 
result eq.~(\ref{gwf}) into a general principle which defines (remnant) chiral symmetry
on the lattice for interacting theories also. Assuming that we are dealing with an action
which is quadratic in the fermion fields, one writes the remnant chiral symmetry condition
as
\begin{equation}
\label{gwi}
\{ h^{-1}_{n_B,n'_B},\gamma^5 \} = 2 \gamma^5 \frac{1}{\kappa_f} R_{n_B,n'_B} \, ,
\end{equation}
where $R$ is a local function.

It has been shown in  \cite{gw} that eq.~(\ref{gwi}) implies the correct 
triangle anomaly on
the lattice. Ginsparg and Wilson made also the remark that all the soft-pion theorems are
expected to be valid if eq.~(\ref{gwi}) is satisfied - I return to this point later.
In the presence of gauge fields however, no $h$ was found in \cite{gw} which was local 
and satisfied eq.~(\ref{gwi}).

The FP QCD action defined by the classical equations eq.~(\ref{fpg}) and eq.~(\ref{ver}) 
satisfies
these conditions. We might consider again a RGT with a very large scale
factor. On the r.h.s. of eq.~(\ref{ver}) the fermion propagates on a very fine lattice over
a very fine gauge field configuration $U$. In this case the fermion propagator goes to
its continuum limit and the only chiral symmetry breaking term will be 
($\omega (U)$ is diagonal in Dirac space)
$1/\kappa_f \delta_{n_B,n'_B}$ as for free fields. 

Multiplying eq.~(\ref{gwi}) by $h$ from left and right one obtains a relation for 
$\{ h,\gamma^5 \}$. Introducing a small bare fermion mass $m_q$ in the action leads to
\begin{eqnarray}
\label{gwim}
\lefteqn{ \{ h(V)_{n_B,n'_B},\gamma^5 \} = \frac{2}{\kappa_f}
(h \gamma^5 R(V) h)_{n_B,n'_B} } \\
& & + 2m_q \gamma^5 M(V)_{n_B,n'_B} , \nonumber
\end{eqnarray}
where $M$ is local. Eq.~(\ref{gwim}) specifies the symmetry breaking terms which enter the
divergence of the axial-vector current \cite{gw}.
One can derive then the following Ward identity in the $m_q \rightarrow 0$ limit:
\begin{eqnarray}
\label{ward}
\lefteqn{ F^2_{\pi} m^2_{\pi}=- \frac{m_q}{2\kappa_f}<(MR)_{0,0}+(RM)_{0,0}>} \\
& - \frac{m_q}{2} < \sum_{n_B} \bar{\chi}_{n_B} M_{n_B,0} \chi_0 
+  \bar{\chi}_0 M_{0,n_B} \chi_{n_B} > &  \nonumber
\end{eqnarray}

The second term on the r.h.s. of eq.~(\ref{ward}) is proportional to the quark
condensate, the first term is the contribution from the remnant symmetry condition.
Both terms go to zero as the bare quark mass $m_q$ goes to zero, and so 
$m_{\pi}\rightarrow 0$. There is no tuning.

Consider the $\sim 1/\kappa_f$ term in eq.~(\ref{gwim}) which enters 
$\partial^{\mu}J^{5a}_{n\mu}$.
The two $h$ factors in this expression will cancel the 2 propagators which connect
this term to other operators in the matrix element considered and produce only a contact
term in Ward identities. 
This is the intuitive reason to expect that the physical consequences
of soft-pion theorems will not be altered, as it is the case in eq.~(\ref{ward}).
This point deserves a detailed investigation.

\section{FP ACTIONS IN 1-LOOP PERTURBATION THEORY}

One can raise formal RG arguments that the FP action is actually 1-loop 
perfect \cite{kw,sol}.
An explicit 1-loop calculation in the $d=2$ non-linear $\sigma$-model seemed to confirm
these arguments: no cut-off effects were found in the finite volume mass gap $m(L)$
using the FP action \cite{we}. 
The same conclusion was obtained in the Schwinger model \cite{fl}.
It has been pointed out recently, however that $m(L)$ is not 
appropriate to illustrate the issue at hand, since for this special quantity the effect
of classical improvement is delayed and shows up first on the one loop level \cite{pis}.

In a recent paper \cite{end} another physical quantity, the smallest energy $E(L,p)$
in the $p \neq 0$ channel was considered, which is free of this accidental nice 
behaviour. The FP predictions for this quantity showed small cut-off effects on 
the 1-loop level.
Therefore, FP actions are {\em not} 1-loop quantum perfect. On the other hand,
the cut-off effect found was $\sim 30$ times smaller with the FP action as that with
the tree level Symanzik improved action.

\section{PARAMETRIZATIONS AND SIMULATIONS}
Solving the FP equations requires patient and skill. The main steps
will be discussed in Sect.~5.2. On the other hand a good parametrization requires 
ideas and insight where there is a lot yet to be learned.

\subsection{Spin and gauge models in $d=2$} 
These investigations show that FP actions, which contain an infinite number of 
couplings, can be excellently parametrized in terms of a limited number of 
operators. The simulation results reflect this way the properties of the FP action
without mixing them up heavily with the errors of poor parametrization.

The general conclusion is the following: the FP action gives very good results
in quantum simulations including questions related to
topology. It performs better than any other improvement schemes tested in $d=2$.

Cut-off effects in the running coupling constant in the $O(3)$ non-linear 
$\sigma$-model are eliminated even at strong coupling where other, more 
straightforward methods fail \cite{hn}. Very small cut-off effects are seen in the 
thermodynamics of the same model \cite{sp}. The parametrizations reproduce the results
expected from exact statements on instantons \cite{top1}. The topological susceptibility
is measured precisely and for the first time consistently 
in the $CP^3$ model \cite{top3}.
In this context let me mention a nice pedagogical work of the MIT group on the
quantum rotor \cite{rot}. This example is also a warning: have more phantasy when 
parametrizing! 

The Schwinger model is the first example where the FP fermion-gauge field
interaction is found and parametrized. The vertex was determined
analytically up to quadratic order in the vector potential \cite{fl}.
The full FP action of the 2-flavour Schwinger model was parametrized and
simulated 
with zero bare quark mass \cite{lp}. At $\beta = 6.$, where the massive excitation
had a mass of 0.33(1) (compared to 0.3257 from continuum theory), the mass
of the 'pion' was found numerically zero (the authors quote 
$m_{\pi} \sim 0.005)$.
This is consistent with the expectations from Sect.~3.5, 
and shows that the 
parametrization is very good down to this $\beta$ value at least.

\subsection{Yang-Mills theory}
Let us discuss first briefly, how to solve the FP equation eq.~(\ref{fpg})
and how to find a parametrization. If $V$ is smooth, $U(V)$ will also be
smooth and the equation can be expanded in terms of the vector potentials.
The couplings entering the quadratic and cubic part of the action
can be determined analytically \cite{sol}. A possible Ansatz for the parametrization
in SU(N) has the form
\begin{eqnarray}
\label{ans}
\lefteqn{ S^{FP}(V)=\frac{1}{N}\sum_C \{c_1(C)[N-ReTr(U_C)]} \\
& & + c_2(C)[N-ReTr(U_C)]^2 + \dots \} , \nonumber
\end{eqnarray}
where the summation goes over a set of loops $C$ one intends to
include in the parametrization. Expanding the Ansatz eq.~(\ref{ans}) in
the vector potentials and comparing it with the quadratic and cubic solution,
the couplings $c_1(C)$ can be fitted. For coarse gauge fields $V$ one has
to solve eq.~(\ref{fpg}) numerically \cite{solp,bn}. The difficulty is that $S^{FP}$
enters on both sides of eq.~(\ref{fpg}). An observation to
proceed is that the minimizing configuration $U(V)$ is much smoother
than $V$ itself. 
Therefore, on the r.h.s. of eq.~(\ref{fpg})
one can use the $\sim c_1(C)$ part of the action, which is already known.

The first parametrizations were rather primitive, they used the trace of 2 
different loops (plaquette and twisted) and their powers \cite{solp,bn}. It turned out 
later that they do not reproduce the theorems on instantons  at all \cite{top2}.
In spite of that the simulated parametrizations performed suprisingly well
on the SU(3) potential, torelon mass \cite{solp,pyl} 
and free energy density $f(T)$ \cite{pap}.
In a contribution to this conference results were reported on the surface 
tension defined at the first order phase transition in $SU(3)$ \cite{papp}. No 
improvement was found. Even more, the results of the parametrized FP action
look rather strange. Further tests are needed here.

Recent works on constructing instanton solutions of the FP action made it
possible to force the parametrization to respect the scale invariance of
these classical objects by adding a third, perimeter-eight loop \cite{kov,papp}. This
opened the way for a study of topological effects in SU(2) \cite{kov}. Rather
than discussing the physical conclusions of this paper let me mention only:
the minimizing configuration $U(V)$, which enters the definition of the
topological charge also, has amazing properties and deserves further study.

\subsection{Fermions, QCD}
Free fermions have been studied in detail including staggered fermions \cite{stag} and
the renormalized trajectory
along the relevant mass direction \cite{wie,hfp1,hfp2,ku}. 
The general conclusion is that the free fermion
FP actions can be parametrized well keeping the couplings on the hypercube only.

It is more difficult to construct the FP form of the fermion-gauge 
interaction. Again, for smooth gauge fields, eq.~(\ref{ver}) can be expanded
in the vector potential and the vertex can be constructed analytically
in the linear order \cite{hfp2,hfpl}. The full FP fermion-gauge interaction has not
been constructed yet.

DeGrand tested several 'FP inspired actions' \cite{dg}. In these actions
the free fermion part has a well parametrized FP form, the 'Pauli term'
resembles to that of the FP vertex, the gauge connections are APE type,
hand made. He found good dispersion relations, (see also in \cite{hfp1})
and reduced cut-off effects.

The author is indebted for the warm hospitality at the Departament 
d'Estructura i Constituents de la Mat\`eria, Barcelona, where this 
paper was completed.

\end{document}